\documentclass[12pt]{article}
 \usepackage{epsfig} \usepackage{amsmath} 
 \usepackage{amssymb} \usepackage{pifont} 
 \usepackage{psfrag} \usepackage{graphicx}
 \newcommand{\ns}{\normalsize}
 \newcommand{\fns}{\footnotesize}

\newcommand{\beq}{\begin{equation}}
\newcommand{\eeq}{\end{equation}} 
\newcommand{\bit}{\begin{itemize}}
\newcommand{\eit}{\end{itemize}} 
\newcommand{\ben}{\begin{enumerate}}
\newcommand{\een}{\end{enumerate}}
 \def\ie{\hbox{\it i.e.}{ }} 
 \def\msbar{\relax\ifmmode\overline{\rm MS}
 \else{$\overline{\rm MS}${ }}\fi}
 \def\eV{\relax\ifmmode{\rm e\kern-0.12em V}
 \else{\rm e\kern-0.12em V{ }}\fi}
 \def\MeV{\relax\ifmmode{\rm M\eV}\else{\rm M\eV}\fi}
 \def\GeV{\relax\ifmmode{\rm G}\eV\else{\rm G\eV}\fi}    
 \def\Acal{\relax\ifmmode{\cal A}\else{${\cal A}${ }}\fi}
 \def\Acalk{\relax\ifmmode{\cal A}_k\else{${\cal A}_k${ }}\fi}
 \def\alps{\relax\ifmmode\alpha_s\else{$\alpha_s${ }}\fi}
 \def\Agoth{\relax\ifmmode{\mathfrak A}
 \else{$\,{\mathfrak A}${ }}\fi}
 \def\tildA{\relax\ifmmode\tilde{A}\else{$\tilde{A}${ }}\fi}
 \def\al{\relax\ifmmode{\alpha}\else{$\alpha${ }}\fi}
 
 \def\Black{}
 
\begin{document} 
 \centerline{\bf\large ``Massive'' 
\ Perturbative QCD, regular in the IR limit}
 \bigskip  
   
 \centerline{\sf\small D.V. Shirkov}\medskip
 
\centerline{\it Bogoliubov Lab., JINR, Dubna, Russia} 
 
 \medskip
  
\begin{flushright} {\it\small Dedicated \ to \ 
   the \ memory \\ of  Petr Stepanovich ISAEV}%
  \end{flushright} 

  \small  \centerline{\bf Abstract}
  
 The goal of research is to devise a modification of 
 the perturbative QCD that should be regular in the 
 low-energy region and could serve as a practical 
 means for the analysis of data below 1 \GeV up to the 
 IR limit. Recent observation of the four-loop pQCD 
 series ``blow-up'' in the region below 1 \GeV for 
 the Bjorken Sum Rule gave an impetus to this attempt.
  
 The proposed \ {\sf ``massive analytic pQCD''} \ is 
 constructed on the two grounds. The first is the 
 pQCD with only one parameter added, the effective \ 
 ``glueball mass'' \ $m_{gl}\lesssim 1\,\GeV\,,$ 
 serving
 as an IR regulator. The second stems out of the 
 ghost-free Analytic Perturbation Theory comprising 
 non-power perturbative expansion that makes it 
 compatible with linear integral transformations.
 
 In short, the proposed MAPT differs from the 
 minimal APT by simple ansatz \ 
 $Q^2 \to Q^2+m_{gl}^2\,.$ 
 \ns 
  
 \section{\large Motivation and Outline} As it is 
 well known, the so-called perturbative QCD (pQCD) 
 or the renorm-group(RG)-improved QCD perturbation 
 expansion {\it taken in the UV limit} is a firmly 
 established part of the particle interaction 
 theory. This piece is not only respectable but 
 worthy of admiration as, starting with 
 gauge-non-invariant quantization, it correlates 
 several dozen of experiments at quite different 
 scales from a few up to hundreds of \GeV.\\ 
 At the same time, the pQCD meets serious troubles 
 in the low energy (large distance) domain below a
 few \GeV at the scales marked by the QCD parameter 
 $\Lambda\sim 400\,\MeV\,.$ This Achilles' heel is 
 related to its UV origin. 
  
 To avoid the unwanted singularity in the LE region, 
 several modifications \cite{modPT} of the pQCD have 
 been proposed. Recently, one of them, the Analytic 
 Perturbation Theory\cite{apt} (APT), was good 
 enough \cite{our2011} in describing the polarized 
 $\Gamma^{p-n}_1(Q^2)=\Gamma_1(Q^2)$ form-factor of 
 the Bjorken Sum Rules (BjSR) amplitude \ {\it 
 down to a few hundred  MeV}. 
 
 The Bjorken moment was presented there as a sum of 
 PT and higher twist (HT) non-perturbative 
 contributions \vspace{-2mm} 
 \beq \label{PT-Bj}                          
 \Gamma_1(Q^2)=\frac{g_A}{6}\biggl[1-
 \Delta^{PT}_{\rm Bj}(Q^2)\biggr]+\Gamma_{HT}\,;
 \quad \Gamma_{HT}= \sum_{i=2}^{\infty}\frac{
 \mu_{2i}}{Q^{2i-2}}\, \,,\eeq \Black
 with $\Delta^{PT}_{\rm Bj}\,$ including the 
 forth-order term $\,\sim(\alps(Q^2))^4\,.$ 
 However, an attempt to fit JLab data by expression 
 (1) with appropriate HT coefficients failed as the 
 perturbative part exploded in the region 0.5 -- 1 
 \GeV and the extracted (via comparison with fitted 
 JLab data) $\mu_{2i}$ values turned out to be 
 unstable w.r.t. higher loop terms in first PT sum. 
 This prevents the description of data below 1 \GeV.
 
 Along with eq.(\ref{PT-Bj}), in \cite{our2011} the 
 PT sum was changed for the APT one 
 \beq\label{Delta_APT-Bj}                     
 \Delta_{\rm Bj}^{PT}=\sum_{k\leq 4}\,\,c_k\,
 (\alps(Q^2))^k\,\quad\Rightarrow \quad
 \Delta_{\rm Bj}^{\rm APT}(Q^2)=\sum_{k\leq 4}\,
 \,c_k\,\Acal_k(Q^2)\,\eeq 
 with $\Acal_k(Q^2)\,,$ the APT ghost-free expansion 
 functions. The positive result consists in good
 fitting of the precise JLab data \ down to a few 
 hundred \MeV with stable HT parameters. \smallskip
 
 This achievement rises hope for the possibility of 
 a global fitting down to the IR limit. Unhappily, 
 none of the above mentioned ghost-free modifications 
 is suitable for this goal. The common drawback is 
 the use of UV logs in the IR region. \medskip
 
 To approach the global fitting of data (like ones 
 for the BjSR form factor), one needs to have a 
 theoretical framework with two essential features:
 \bit\item \ Correspondence with common pQCD in the 
 UV (that is above a few GeV);
 \item \ Correlation with lattice simulation results 
 for the effective coupling $\alps(Q)\,$ smooth 
 behavior in the low-energy domain.\eit 
 
 As a primary launch pad for this construction, the 
 above mentioned APT seems good. It satisfies the 
 first of the conditions and, qualitatively, the 
 second one. To exempt the APT-like scheme from its 
 last drawback -- the singularity (with an infinite 
 derivative) in the IR limit, one has to disentangle 
 it from the UV logs. To this goal, a mass-dependent 
 RG-invariant modification inspired by our paper 
 \cite{Sh99-Un} will be used. 
 
 In Sect.3, on the basis of the massive renorm-group 
 (see, Sect.2), the non-singular version of pQCD 
 with one additional (besides $\Lambda$)\, parameter, 
 an effective ``gluonic'' mass, \ $m_{gl}\,,$ a {\it 
 massive pQCD} \ -- MPT for short, is formulated.

 \section{\large Massive Perturbation Theory    
 \label{mpt}} In \cite{Sh99-Un}, a particular way 
 of constructing the QCD invariant coupling 
 $\alps(Q^2)\,$ free of unphysical singularities 
 was proposed. In contrary to the APT, it does not 
 involve explicit nonperturbative contributions. 
 Instead, the $\,Q^2\,$ algebraic (non-log) dependence 
 appeared there due to threshold effects, and an 
 essential technical ingredient was the assumption 
 of the {\it finite gluon mass} \ formal presence.

 The model expression for $\,\alps(Q^2)\,$ was 
 obtained there by the RG summation of the \ {\sf 
 mass--dependent} \ diagram contribution -- see 
 below Eqs. (\ref{a1mrg}) and (\ref{al2mrg}). 
 It depends upon the gluon $\,m_{\rm gl}\,$ and 
 light quark $\,m_{\rm lq}\,\,(lq= u,d,s)\,$ masses; 
 in the IR region $Q^2>0\,$ has no singularities  
 with a finite limiting value $\,\alps(0)\,$ and, 
 as $\,Q^2/m^2 \to\infty\,,$ smoothly transits into 
 the usual asymptotic freedom formula.\vspace{-3mm}
 
 \subsection{\ns Mass-dependent 1-loop diagram}
 At the one-loop level, the starting element is the 
 massive (mass-dependent) 1-loop contribution. For 
 example, to the virtual dissociation of a vector 
 particle (photon, gluon) into a massive fermion 
 antifermion pair ($e^++e^-$\,;\, $\,q+\bar{q}\,$) 
 in the $s$--wave state, there corresponds a 
 function $\,I_s(Q^2/m^2)\,$ representable via 
 spectral integral\footnote{For an explicit 
 expression see Sect.24.1 in the text-book \cite{QF} 
 and Sect.35.1 in the monograph \cite{kniga}.} 
 \vspace{-4mm}  

 {\small\beq\label{fot-polariz}             
 I_s(z)=z\int\limits_{1}^{\infty}\tfrac{k_s(\sigma)
 \,d\,\sigma}{\sigma(\sigma+z)};\quad k_s(\sigma)=
 \sqrt{\tfrac{\sigma-1}{\sigma}}\,;\quad I_s(0)=
 0.492\,,\quad I_s^{'}(0)=2/3\,.\eeq}
 
 \noindent which in the space-like region $\,z>0\,$ 
 is a \ {\it positive, monotonically growing 
 function} \ with the log asymptotic behavior  
 $I_s^{UV}(z=Q^2/m^2)\,\simeq\,\ln z-C_s+O(1/z)\,;
 \,\, C_s=2(1-\ln 2)\,$ and the regular IR limit 
 with a finite derivative.

 \subsection{\ns Massive Renorm--group summation}
 For the QCD coupling modification at small 
 space-like $\,Q^2\lesssim\Lambda^2\,$ we involve 
 the mass-dependent Bogoliubov renorm-group (mRG) \ 
 formulated in the pioneer RG papers \cite{dan55} 
 in the mid-fifties. As it is known since that time 
 \cite{blank}, the mRG, like the common massless RG,
 sums iterations of a one-loop contribution\footnote{
 \Black Here and below, the superscript in square 
 brackets $\Acal^{[\ell]}$ denotes the order of 
 loop approximation.} \ $\alps(z)_{\rm pt}^{[1]}=
 \alps-\alps^2\,A_1(z,y)+\dots\,,$ into the 
 geometric progression\vspace{-5mm}
   
 \beq\label{a1mrg} \alps(Q^2)_{\rm rg}^{[1]}=
 \frac{\alps}{1+\alps A_1(z,y)}\,;\quad z=\frac{
 Q^2}{m^2}\,,\quad y=\frac{\mu^2}{m^2}\,.\eeq 
  For the 2-loop case, with $A_2\,,$ the genuine 
 second-loop contribution \vspace{-2mm}
  
 \beq\label{pert2}  \phantom{WWWWW}
 \alps(z,y)_{\rm pt}^{[2]}=\alps-\alps^2 A_1(z,y)+
 \alps^3\left(A_1^2- A_2(z,y)\right)+\dots\,,\eeq
 an analogous (approximate) \ RG-invariant \ 
 ``massive''  \ sum  
 \beq \alps(z,y)_{\rm rg}^{[2]}=\frac{\alps}   
 {1+\alps\,A_1(z,y)+\alps\frac{A_2(...)}{A_1(...)}\,
 (1+\alps A_1(...)) }\;.\label{al2mrg}\eeq 
 was also devised later (eq.(8) of paper
 \cite{mass92}). There, 
 \beq\label{eq7}                              
 \,A^{[\ell=1,2]}(z,y)=I^{\ell}(Q^2/m^2)-I^{\ell}
 (\mu^2/m^2)|_{UV}\quad\to\quad\beta_{\ell-1}\,
 \ln(Q^2/\mu^2)\,,\eeq 
 with transition to the QCD scale\vspace{-6mm}
 
 \beq\label{8}\phantom{WWWWWWWW}
 \alps(Q^2)^{[1]}_{\rm rg}
 =\frac1{\beta_0\,\ln(Q^2/\Lambda^2)}\,\eeq
 performed via the relation\vspace{-4mm}
 
 \beq\label{9}\phantom{WWWW}1/\beta_0\,\alps +
 \ln(Q^2/\mu^2)=\ln(Q^2/\Lambda^2)\,.\eeq
 
 \section{\large The MPT construction
  \label{mpt-constr}}
  \subsection{\ns One-parameter massive Model}  
 A simple idea is to change the usual UV logarithm
 $\ln x \ (x=Q^2/\Lambda^2)\,$ that is singular also 
 in the IR for the \ ``long logarithm'' $L_{\xi}(x)
 =\ln(\xi+x)\,$ which reproduces qualitatively the
 smooth LE behavior of function (\ref{fot-polariz}) 
 being regular at $\,Q^2=0\,.$ It is noteworthy that 
 the new parameter is expressed via the coupling 
 constant (at $Q^2=0$) by the relation \ $\xi=
 e^{1/\beta_0\,\alps}\,$ non-analytic at $\alps=
 0\,.$ It corresponds to  the ``effective 
 gluonic mass'' \ $m_{gl}=\sqrt{\xi}\,\Lambda\,,$ 
 an old notion (see, a recent survey by Simonov 
 Ref.\cite{simon10} and references therein) used 
 as an IR regulator. In short, our ansatz is \\
 \phantom{WWWWWWWWWWWAAAAAB}\framebox{$Q^2 \to 
 Q^2 + m_{gl}^2\,.$}\hspace{56mm} (A)\medskip
 
 {\sf One-loop case.} \ The \ ``1-loop 
 structure'' \ in the denominator of eq.(\ref{8}) 
 is changed now for the \ ``long logarithm'' \ with 
 $\,\xi\,,$ an adjustable parameter \vspace{-4mm}
                    
 \beq\label{10} \ln x \to L_{\xi}(x)=            
 \ln(\xi+x)\,,\quad x=Q^2/\Lambda^2\,.\eeq  
 At moderate LE scales the form \vspace{-4mm}
 
 \beq                                            
 L_{\xi}(x)=\ln\xi+\ln(1+\phi\,x)=\tfrac1{\beta_0
 \,\alps}+\ln(1+\phi\,x)\,;\quad\phi=1/\xi\,\eeq 
 is more adequate. In terms of this LE form, one 
 has  \beq\label{a1-mpt}                        
 \Acal_{1 MPT}^{[1]}(x; \xi)=\frac1{\beta_0\,
 \ln(\xi+x)}=\frac{\alps}{1+\alps\,\beta_0\,
 \ln(1+x\,\phi)};\quad\phi=e^{-1/\beta_0\,\alps}
 \,,\eeq
 with $\,\alps=\alps(x=0);\,\,\,\alps|_{\xi=10\pm2}
 =0.61\mp0.05\,.$ The finite derivative at $\,Q=
 0\,$ also is of interest $\,\Acal_{1,MPT}^{[1]'}
 (0,\xi)=-\beta_0\,\alps^2\,.$ \bigskip
   
 {\sf The 2-loop 1-parameter model.}  \     
 Starting with the 2-loop massive RG-summed result, 
 eq.(\ref{al2mrg}) corresponding to the 
 mass-dependent PT expansion, eq.(\ref{pert2}), we 
 generalize eq.(\ref{a1-mpt}) by using \ {\it the  
 same \ ``long  logarithm'' \ model  for  the  
 second-loop  contribution} \vspace{-4mm}
 
 \beq\label{A21mass}                          
 A_2(x\,\phi)=\beta_1\,\ln(1+\phi\,x)\,.\eeq
  \vspace{-5mm}%

 \noindent That is \vspace{-4mm} {\small  
 \beq\label{2l-MPT}\phantom{AA}               
 \Acal^{[2]}_{1,MPT}(x,\xi)=\frac{\alps}{1+\alps\,
 \beta_0\,\ln(1+\phi\,x)+\alps\,\frac{\beta_1}
 {\beta_0}\,\ln[1+\alps\,\beta_0\,\ln(1+\phi\,x)]
 \,.} \eeq }
 
 Now, the condition \ $\alps(M_{\tau}^2)=0.34\,\,$ 
 can be used for a rough evaluation of an \ 
 ``effective MPT-QCD scale'' \ value. At the NLO 
 case, the $\Lambda^{[2]}(\xi)\,$ dependence on 
 $\xi$ turns out to be rather weak \footnote{with 
 values less than the pQCD one \ $\Lambda^{n_f=3}
 =420\pm 10\,\MeV\,.$} :
 \beq\label{lambda}                             
 \Lambda^{[2]}(10\pm2)\sim 315\mp10\,\MeV\,.\eeq 
 The related $m_{gl}=\sqrt{\xi}\,\Lambda^{[2]}\,$ 
 value could be close to the nucleon mass at 
 $\,\xi\sim10\,.$ 

 \subsection{\ns Higher MPT expansion functions} 
 In the construction under devising, we intend to
 preserve an essential APT feature, namely, the 
 non-polynomiality of the modified ``perturbative''
 MPT-expansion, expansion over a set\footnote{The same 
 symbol $\Acal,$ as in the minimal APT with limiting 
 relation $\Acalk(x,\xi=0)=\Acalk(x)$ is used.} of 
 functions $\left\{\Acal_k(Q^2,\xi)\right\}$
 connected by the same differential relations as in 
 the APT (with the dotted notation for logarithmic 
 derivative  $\dot{F(x)}=x\,F^{'}(x)$)
 \beq\label{recurr2}                         
 -\tfrac xk\,\tfrac{\partial}{\partial x}\Acalk(x,\xi)
 :=-\tfrac1k\,\dot{\Acal}_k(x,\xi)=\beta_0\,\Acal_{k+1}
 (x,\xi)+\beta_1\,\Acal_{k+2}(x,\xi)+\dots \,\,.\eeq
  To the arguments ascending to the 80s 
 \cite{pi-terms} and related to the $\pi^2$-terms 
 summation procedure in the s-channel (see, also 
 Refs.\cite{apt}), one can add a fresher reasoning 
 \cite{non-polynom}. \smallskip  
 
 {\sf The second MPT function.} This recurrence 
 property ensures compatibility \cite{dv01} with 
 linear transformations involved in transition to the 
 distance picture (Fourier-conjugated with the 
 momentum-transfer one) and to the annihilation 
 s-channel. 
 
  In a particular case $k=1$, with (\ref{a1-mpt}), 
 and neglecting the second r.h.s. term, \ie using 
 the one-loop relation
 \beq\label{recurr1}                             
 \Acal_{k+1}(x,\xi)= -\tfrac1{k\,\beta_0}\,
 \dot{\Acalk\,(x,\xi)} \eeq
 as a definition for higher functions, one gets 
 the second expansion function\footnote{For 
 discussion of a more accurate definition of the 
 two-loop second function, see Appendix A.}\vspace{-3mm}
 
 \beq\label{a2mpt}                              
 \Acal_2(x,\xi)=\left(\Acal_1(x,\xi)\right)^2\,
 R(x,\xi)\,;\quad R(x)=\dot{L}_{\xi}(x)=\frac{\phi
 \,x}{1+\phi\,x} \,;\quad \phi=1/\xi\,,\eeq
 that turns to zero in the IR limit. Besides, \
 $\Acal^{'}_2(0,\xi)=\alps^2\,\phi\,.$\medskip 
  
 {\sf The third MPT expansion function} obtained 
 by  eqs.(\ref{recurr1}) and (\ref{a2mpt}) can
 be represented in the form \vspace{-4mm}
 \beq\label{a3mpt} \phantom{AAA}                 
 -2\,\beta_0\,\Acal^{[1]}_{3,MPT}(x,\xi)= 
 \dot{\Acal}_{2,MPT}(x,\xi)=2\,\Acal_{1,MPT}\,
 \dot{\Acal}_{1,MPT}\,R(x)+(\Acal_{1,MPT})^2\,
 \dot{R}(x)\,, \eeq
 that is sufficient for perceiving IR 
 properties $\,\Acal^{[1]}_{3}(0,\xi)= 0 ;\,\,
 \Acal^{[1] '}_{3}(0,\xi)=-\alps^2\,\phi\,.$%
 
 \section{\large General features of the MPT 
 scheme} Here, we shortly discuss some important
 features of the proposed MPT construction.
 \subsection{\ns Annihilation channel}  For the 
 transition to the s-channel, $f(Q^2)\to F(s)$
 one uses the spectral representation and the 
 Adler-type relation \vspace{-4mm}
 \[\phantom{WWWWWW}f(Q^2)=\tfrac1{\pi}\,
 \int_0^\infty\,\frac{\rho(\sigma)\,d\sigma}{Q^2
 +\sigma}\,,\quad f(Q^2)=Q^2\int_0^\infty\,
 \frac{F(s)\,ds}{(s+Q^2)^2}\,,\]
 which, in turn, results in \vspace{-7mm}
 
 \beq\label{A_n-Mink-rho} \phantom{MMMMMM}      
 F(s)=\frac{1}{\pi}\int_s^\infty\,\frac{d\sigma}
 {\sigma}\,\rho(\sigma)\,;\quad \rho(\sigma)=
 \Im f(-\sigma-i\,\varepsilon)\,.\eeq
 
 In the simplest one-loop case (\ref{a1-mpt}),
 \beq\label{21}                                
 \beta_0\,\Acal_{1 MPT}^{[1]}(x; \xi)=\frac1{
 \ln(\xi+x)}=\tfrac1{\pi}\,\int_0^\infty\,
 \frac{\rho_1(\sigma)\,d\sigma}{x+\sigma}\,,
 \quad \rho_1(\sigma)=\frac{\pi\,\theta(\sigma-
 \xi)}{\ln^2(\sigma-\xi)+\pi^2}\,,\eeq
 \beq\label{22}                                
 \beta_0\,\Agoth_{1 MPT}^{[1]}(s;\xi)=\frac{1}{\pi}
 \int_s^\infty\,\frac{d\sigma}{\sigma}\,
 \rho_1(\sigma)\,=\frac1{\pi}
 {\arctan\frac1{L_-(s)}}\,;\quad L_-(s)=\ln\left(
 \frac{s-m_{gl}^2}{\Lambda^2}\right)\,.\eeq
 This expression remains regular around 
 $\,s\sim m_{ gl}^2 + \Lambda^2\,$ and constant 
 below $\,s= m_{ gl}^2\,.$ 
 
 To obtain the higher $\Agoth_{k MPT}$ functions,  
 one should use a differential recurrent relation, 
 just as in APT. For the two-loop case, one can 
 combine it with the {\it two-loop effective log 
 $L^*$}, a trick proposed in Ref.\cite{ssh-99} and 
 further developed in Ref.\cite{trick2}.

 \section{\large Comparison with APT and Discussion}
 To compare the new construction with the APT one, 
 in  Fig.1 we give the curves at a few values of 
 $\,\xi=10\pm2\,$ in the region below 2 \GeV for 
 the first MPT function $\Acal^{[2]}_{1,MPT}\,$ 
 (\ref{2l-MPT}) together with the corresponding 
 APT (dashed) curve.  \vspace{-6mm}
 
 \begin{figure}[h]\begin{center}
 \centering{\epsfig{scale=0.65,figure=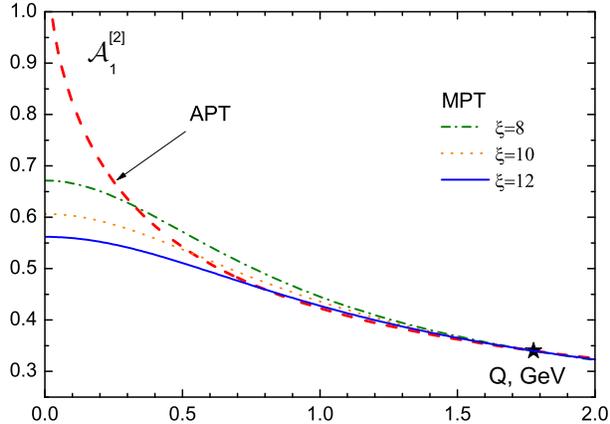}}
 \vspace{-8mm}
 \caption{\fns The first MPT function (for a few 
 $\xi$ values) in comparison with the APT one.}
  \label{mptNLO-A1} \end{center} \end{figure}
 
  It can be seen from the NLO curves that values 
 $\xi=8-10\,$ seem to be preferable. Indeed, for 
 these values, the first MPT function,
 $\Acal^{[2]}_1(x,\xi)$ is reasonably close to the 
 first APT one down to 1 \GeV. At the same time, in 
 the region around 500-700 \MeV it deviates from APT 
 but is more similar to the results of lattice 
 simulations, especially to the Orsay 
 group\cite{orsay,latt-rev} ones. \vspace{-5mm}
   
\begin{figure}[h]\begin{center}
 \epsfig{scale=0.65,figure=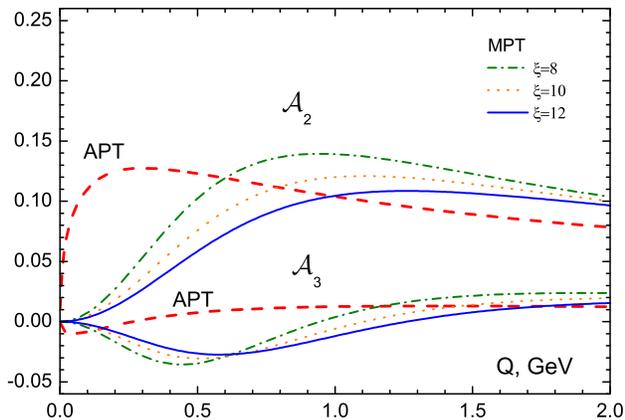}\\
  \vspace{-10mm}
  
 \caption{\fns The second and third MPT functions 
 vs. the APT ones.} 
  \label{mptNLO-2+3} \end{center} \end{figure}
  Figure 2 exposes the second and third MPT 
 functions as roughly estimated by the one-loop 
 Eqs.(\ref{a2mpt}),(\ref{a3mpt}). 
 It is seen that instead of singular IR slopes of 
 all APT thick dashed (red) curves, all the MPT 
 functions are the IR regular ones. At the same 
 time, just as in APT, the second and third MPT 
 functions $\,\Acal_{2,3; MPT}\,$ are noticeably 
 smaller than square $(\Acal_{1,MPT})^2$ and cube 
 $(\Acal_{1,MPT})^3 $ of the first one. E.g., at 
 $\,Q=0.5\,$ \GeV and $\,Q=1.0\,$ \GeV the reduction 
 factors for the $\,\Acal_{3; MPT}\,$ are about 0.2 
 and between 0.1 - 0.2 correspondingly. Besides, the 
 $\Acal_{3,MPT}$ is negative below 1 GeV. \smallskip  
 
  Now, the \ ``MPT-perturbative'' \ expansion 
 similar to eq.(\ref{Delta_APT-Bj}), due to recurrent
 relation (\ref{recurr1}) can be represented in a 
 form of Taylor series expanded over the parameter
 $\vartriangle\ln x=c_1\,\pi/\beta_0=1.60\,$ with 
 the final effect \vspace{-2mm}
 \beq\label{23}
 \Delta_{MPT}^{*}\simeq\tfrac1{\pi}\,\Acal_{1,MPT}
 (x^*)\,;\quad x^*=x\,e^{-\vartriangle\ln x}=Q^2/
 (\Lambda^*)^2\,,\quad\Lambda^*=2.25\,\Lambda\,.\eeq
  One should keep in mind that both the logarithm 
 shift $\vartriangle\ln x\,$ and the ``one-term 
 approximation'' (\ref{23}) error
 $\,\delta\Delta_{MPT}\,\sim\Acal_{3,MPT}\,$ are
 scheme-dependent quantities. In the \msbar scheme
 under consideration, $\,\delta\Delta_{MPT}\,$
  -- due to the smallness of the reduction-factor 
  -- is negligible. 
  
  However, one can get another angle on  
 Eq.(\ref{23}) and return to the old idea of the 
 {\it effective coupling constant} \cite{ecc} which 
 is not so far from \ ``RESIPE'' \cite{resipe} and 
 from  the \ ``commensurate scale relations''
 \cite{csr} concepts. Then, the new scale can be 
 treated as a specific one for the given process; 
 $\,\Lambda^*=\Lambda_{Bj}\,.$  \medskip
 
 \centerline{\sf Table. \ ``Glueball mass'', and 
 $\Lambda_{Bj}$ \ for a few values of $\,\xi.$}
 \vspace{-3mm} 
  \begin{table}[!h] 
 \begin{center}\label{tab:Bjtot_PT}
 \begin{tabular}{|l||c|c||c|c||c|} \hline
 $\xi$&$\Lambda_1$&$m_{gl}^{[1]}$&$\Lambda_2$
 & $m_{gl}^{[2]}$ & $\Lambda_{Bj}$ \\ \hline 
 8 & 244 &690 &324 & 915  & 730 \\
 10& 249 &787 &315 & 995  & 710 \\
 12& 253 &876 &305 & 1160 & 686 \\ \hline
\end{tabular} \end{center}
\end{table} \vspace{-4mm} 
  The ``glueball mass'' values given in the Table 
 for the LO and NLO cases also look attractive. 
 They can be confronted with the glueball mass 
 $M^{Q\bar{Q}}_{2g}\sim 1-2\,\GeV$ of paper 
 \cite{simon10} and with gluon mass $M\sim 500$ 
 \MeV from the lattice estimate \cite{orsay} as 
 well as from solution of the Schwinger-Dyson eqs. 
 (see Ref.\cite{binosi12} and references therein).
 \medskip  
 
 Besides, as it can be shown \cite{KhSh12}, the MPT 
 perturbative sum $\Delta_{MPT}$ together with a 
 duly modified HT sum allows one to fit the JLab 
 data down to the very IR limit -- see below Fig.4 
 in the Appendix B.
 There, the generic HT function was conjectured in 
 the IR-regular form $\,\mu_4\,(Q^2+m_{ht}^2)^{-1}\,$ 
 with the only parameter. It is remarkable that its 
 value $m_{ht}\sim 0.7-1\,\GeV\,$ is close to the 
 $m_{gl}$ \ one. This gives hope that ansatz (A) 
 reflects some general physical essence. \medskip 
 
 \centerline{\bf\ns Acknowledgements}\medskip 
 It is a pleasure to thank Oleg Teryaev for 
 stimulating discourses, Michael Ilgenfritz and 
 Andrej Kataev for discussion as well as 
 Vjacheslav Khandramai for useful advice and 
 technical help. This research has been partially 
 supported by the Presidential grants for support 
 of Scientific School 3810.2010.2, 3802.2012.2 
 and by RFFI grant 11-01-00182.
  
 \bigskip 

  \centerline{\sf\large Appendix A : \ 2-loop MPT 
 higher functions}\medskip\small
       
  For a more accurate definition of the 2-loop 
 higher expansion functions, one could use 
 recurrent relation (\ref{recurr2}) at $k=1$ and 
 truncated eq.(\ref{recurr1}) for the $k=2$ case. 
 
 With the technical notation \
 $\varphi(t=\ln x)=-(1/\beta_0)\,\dot{\Acal}_{
 1,MPT}(x)\,;\,\,\Acal^{[2]}_{2,MPT}(x)=y(t)\,$
  one gets two relations of eq.(\ref{recurr2}).
 Neglecting $\Acal_{4,MPT}$ we come to boundary
 value problem 
 \beq\label{diff-eq-1} y(t)-\theta\,\dot{y}(t)=
 \varphi(t)\,,\quad y(\infty)=0\,;\qquad\theta=
 \tfrac{\beta_1}{2\,\beta_0^2}\eeq
 and auxiliary relations \vspace{-2mm}
 \beq \phantom{WWWWWWW}                     
 \Acal_{3,MPT}^{[1]}(x,\xi)=-\tfrac1{2\,\beta_0}
 \,\dot{\varphi}(t)\,,\quad\Acal_{4,MPT}(x,\xi)=
 \tfrac1{6\,\beta_0^2}\,\ddot{\varphi}(t)\,,\eeq
 Solution of (\ref{diff-eq-1}) \  $y(t)=
 \int^{\infty}_0\,e^{-s}\,\varphi(t+s\,\theta)
 \,d\,s $ \ being expanded in powers of $\theta$ 
 yields the form 
  \beq\label{solA2exp}                      
 \Acal^{[2]}_{2,MPT}(x,\xi)=-\tfrac1{\beta_0}\,
 \dot{\Acal}^{[2]}_{1,MPT}(x,\xi)-\tfrac{\beta_1}
 {\beta_0}\,\Acal_{3,MPT}+O(\Acal_{4,MPT})\,,\eeq
 completely correlating with eq.(\ref{recurr2}).
 
  On the other hand, one can use an approximate, 
 `two-loop effective log trick'' of papers 
 Ref.\cite{trick2} \vspace{-2mm}
 \beq\label{28}\phantom{WWWW}\ell=\ln x     
 \to\mathcal{L}_2[\ell]:=\ell+b\,\ln\sqrt{\ell^2+
 2\pi^2}\,,\quad b=\beta_1/\beta_0\,.\eeq  
 
  Combining this with eqs.(8) and (10) one gets  
 (with $L_{\xi}(x)=\ln(\xi+x)$)
 \beq\label{28}                            
 \Acal_{1 ,\ell_2}^{[2]}(x,\xi)=\frac1{\beta_0\,
 \mathcal{L}_2[L_{\xi}(x,\xi)]}\,,\quad \Acal_{2,
 \ell_2}^{[2]}(x,\xi)=\frac{\mathcal{R}(x)}{\beta_0^2
 \,(\mathcal{L}_2[L_{\xi}(x)])^2}\,,\quad
  ... \,\, ;\eeq 
 \[\mathcal{R}(x)=\mathcal{L}^{'}_2[L_{\xi}(x)]\,
 R(x)=\left(1 + b\frac{L_{\xi}(x)}{L_{\xi}^2(x)+2
 \pi^2}\right)\,R(x),\quad R(x)=\frac x{\xi+x}\,.\] 
  
  With due account for the numerical values 
 $\,\beta_0(n_f=3)=0.716\,;$ $\,b=0.566\,,$
 one has \vspace{-2mm}
 
 \[\beta_0\,\alps=\frac1{\ln\xi+b\,\ln\sqrt{(
 \ln\xi)^2+2\pi^2}}\,;\quad \Lambda_{2,\ell_2}
 =\frac{1.777}{\sqrt{23.33-\xi}}\,.\]
 with\footnote{For the practical use of the last 
 simple relation see Ref.\cite{pst10}.} \ \ 
 $\alps|_{\xi=10\pm2}=0.435\mp0.03\,;\quad 
 \Lambda_{2,\ell_2}(\xi=10\pm2)=490\pm35\,\MeV\,
 \sim  1.95\,\Lambda_1\,. $ 
 \begin{table}[!h] 
 \begin{center}\label{tab:Bjtot_PT}
 \begin{tabular}{|l||c|c||c|c||c|c|} \hline
 $\xi$&$\Lambda_1$&$m_{gl}^{[1]}$&$\Lambda_2$& 
 $m_{gl}^{[2]}$&\alps&$\Lambda_{2,\ell_2}$\\ \hline 
 8 & 244 &690 &324 & 915 &0.438 & 455\\
 10& 249 &787 &315 & 995 &0.435 & 490\\
 12& 253 &876 &305 & 1160 & 0.432 & 525 \\ \hline
\end{tabular} \end{center}
\end{table}  \vspace{-2mm} 
 
    These expressions can be confronted with the
 previous ones eqs.(\ref{2l-MPT}),(\ref{a2mpt}). 
 For example, at $Q\sim500\,\MeV; x\sim1\,$ and
 $\Acal^{[2]}_{1,\ell_2}(1,\xi=8)\simeq0.45\,,\quad
 \Acal^{[2]}_{1,\ell_2}(1,\xi=12)\simeq0.40\,.$ 
 
  In the context of relation (\ref{23}), the 
 second term in the r.h.s. of the last expansion 
 (\ref{solA2exp}) reduces further the error 
 $\delta\Delta^{*}\,$ of expression (\ref{23}) 
 for $\,\Delta^{*}_{MPT}\,.$ \bigskip \newpage
 
 \centerline{\sf\large Appendix B : \ The Ansatz 
 (A) effect on the Bjorken Sum Rule analysis}
 \medskip\small 
  
 The net effect of the Ansatz (A) used literally 
 (but roughly) can be described as a transition to 
 the new momentum-transfer scale 
 in both perturbative (PT) and higher-twist
 (HT) items. Explicitly, in Eq.(1), this means
 \beq\label{mpt}\Delta^{PT}_{\rm Bj}(Q^2)\to 
 \Delta^{MPT}\sim\Delta^{PT}_{\rm Bj}(Q^2+m_{gl}^2)
 ;\quad \Gamma_{HT} =\frac{\mu_4}{Q^2}
 \to G_{HT}= \frac{\mu_4}{Q^2+m^2_{ht}}\,. \eeq
 
 Meanwhile, as it was shown above, under a more 
 detailed analysis (that includes differential 
 recurrent relations) the correspondence is more 
 intricate -- see, e.g., Figs 1 and 2. 
 
 Nevertheless, it is evident by observation that 
 for $m_{gl}\sim 500\,$\MeV the solid (green) 
 curve from Fig.3 (taken from paper \cite{our2011}) 
 visually corresponds to Fig.4 curve (according 
 to \cite{KhSh12}) with ``shifted'' 
 scale $Q_{\xi}^2=Q^2-m^2_{gl}$. 
   \begin{center}
\begin{tabular}{lr}
 \begin{minipage}{70mm} \vspace{-12mm}
 \begin{center}
  $$\includegraphics[width=0.9\textwidth]
 {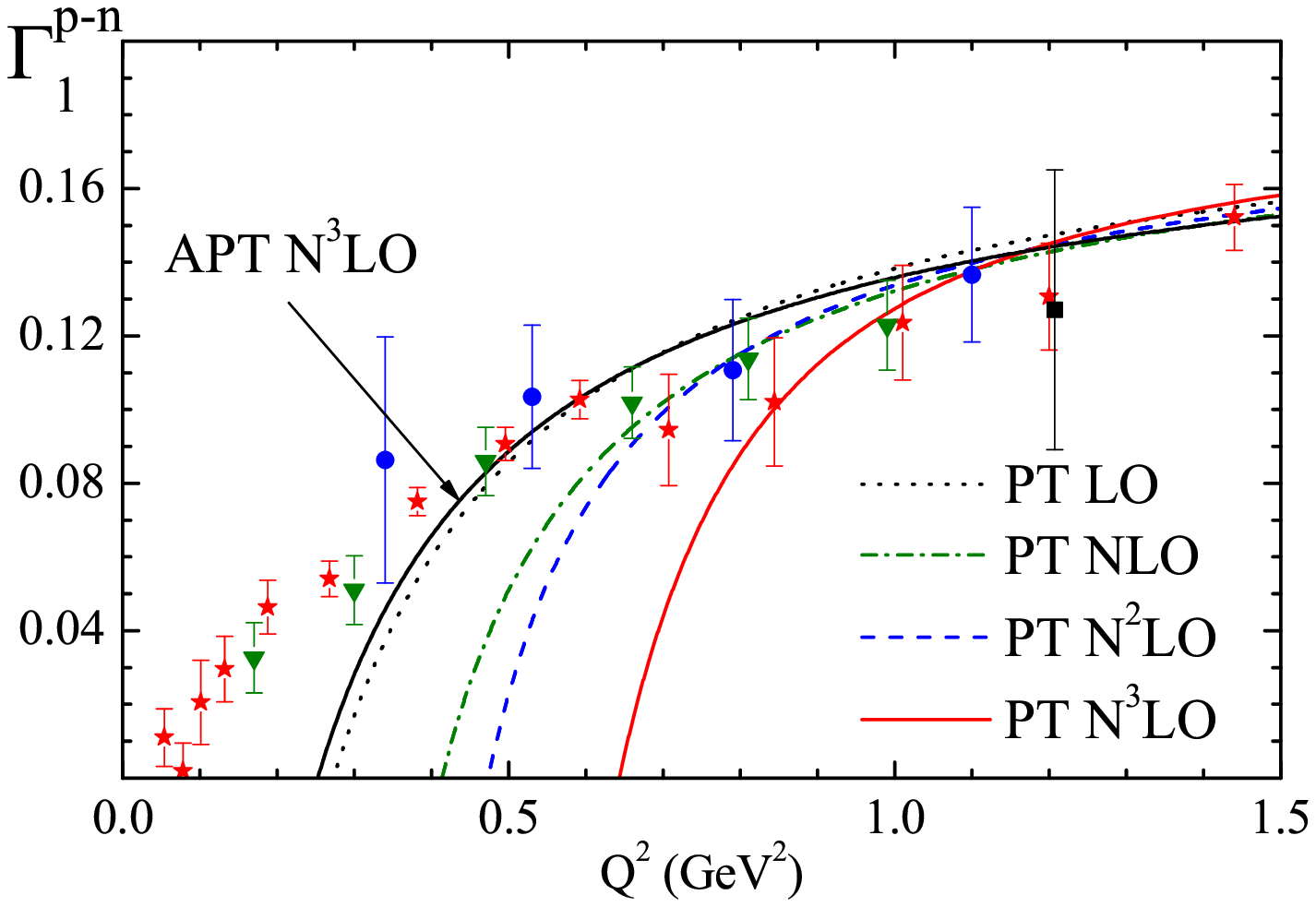}$$  \vspace{-6mm}
  {\small Fig.3 : Figure 5 from paper \cite{our2011}}
  \end{center} \end{minipage}
 
 &\hspace{-4mm}\begin{minipage}{75mm}
\begin{center} \vspace{-5mm}
 $$\includegraphics[width=0.9\textwidth]
 {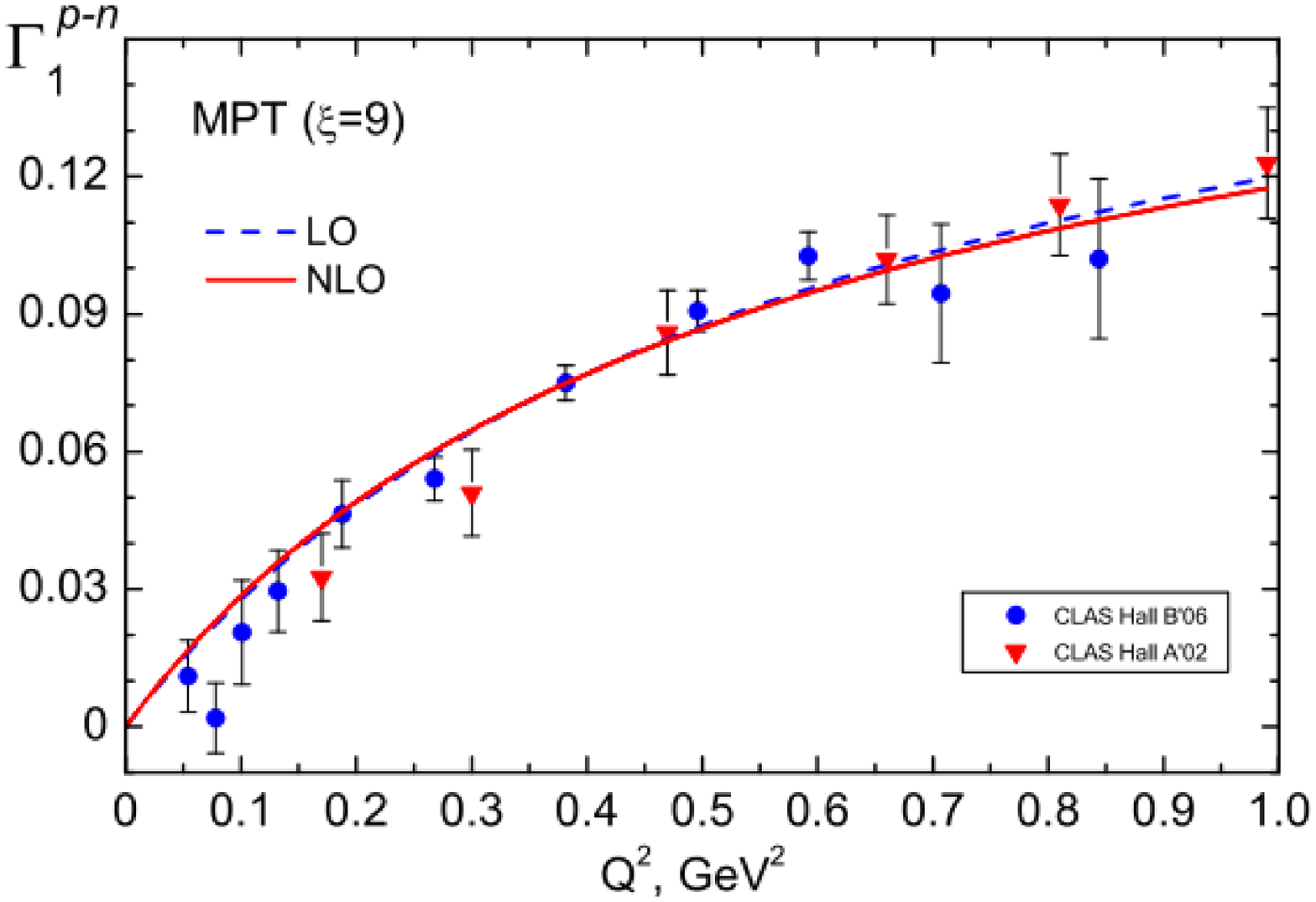}$$ \vspace{-9mm}
 
 {\small Fig.4 : The MPT fitting of the JLab data 
 with change (\ref{mpt}) used, according to 
 \cite{KhSh12}.}
 \end{center} \end{minipage}
 \end{tabular} \end{center} 
  
  \end{document}